\documentclass[10pt, fullpage]{article}

\usepackage{float, graphicx}
\usepackage{amsfonts, amssymb, amscd}
\usepackage{amsmath, delarray, theorem}
\usepackage{natbib}
\bibpunct{(}{)}{;}{a}{}{,}
\renewcommand{\cite}{\citep}

\usepackage{myfile} 
\usepackage{url} 

\newcommand{\btab}{\begin{table}[H]
        \begin{center}}
\newcommand{\etab}{\end{center}
        \end{table}}

\newcommand{\bfig}{\begin{figure}[H]}
\newcommand{\efig}{\end{figure}}
\newcommand{\mycap}[1]{\small \caption{#1}}
\newcommand{\bitem}{\begin{itemize}}
\newcommand{\eitem}{\end{itemize}}
\newcommand{\benum}{\begin{enumerate}}
\newcommand{\eenum}{\end{enumerate}}
\newcommand{\beqnn}{\begin{eqnarray*}}
\newcommand{\eeqnn}{\end{eqnarray*}}
\newcommand{\beqn}{\begin{eqnarray}}
\newcommand{\eeqn}{\end{eqnarray}}

\newcommand{\myvec}[1]{\mathbf{#1}}

\newcommand{\bfourmatrix}{\begin{array}({cccc})}
\newcommand{\efourmatrix}{\end{array}}
\newcommand{\bvector}{\begin{array}({c})}
\newcommand{\evector}{\end{array}}

\newcommand{\gvec}[1]{\mbox{\boldmath ${#1}$}}
\newcommand{\dsum}{\displaystyle\sum}

{\theoremstyle{plain}
  }
{\theoremstyle{plain}
  }
{\theoremstyle{plain}
  }
{\theorembodyfont{\rmfamily}
  }
{\theorembodyfont{\rmfamily}
  }

\begin{document}

\def\mytitle{%
Kernels and Ensembles: Perspectives on Statistical Learning}

\def\myabstract{%
Since their emergence in the 1990's, the support vector machine and the 
AdaBoost algorithm have spawned a wave of research in statistical machine 
learning. Much of this new research falls into one of two broad 
categories: kernel methods and ensemble methods. In this expository 
article, I discuss the main ideas behind these two types of methods, 
namely how to transform linear algorithms into nonlinear ones by using 
kernel functions, and how to make predictions with an ensemble or a 
collection of models rather than a single model. I also share my personal 
perspectives on how these ideas have influenced and shaped my own 
research. In particular, I present two recent algorithms that I have 
invented with my collaborators: LAGO, a fast kernel algorithm for 
unbalanced classification and rare target detection; and Darwinian 
evolution in parallel universes, an ensemble method for variable 
selection. }

\def\mykeywords{
AdaBoost; 
kernel PCA;
LAGO;
parallel evolution;
random forest;
SVM.}

\title{\mytitle}
\author{Mu Zhu, PhD\\
University of Waterloo \\
Waterloo, Ontario, Canada N2L 3G1}
\date{\today}
\maketitle

\begin{center} 
{\bf\large Abstract}
\end{center}  

\vspace{0.5cm}

{\small \myabstract}

\vspace{0.5cm}


%\tableofcontents

\section{Introduction}
\label{sec:intro}

The 1990's saw two major advances in machine learning: the support vector 
machine (SVM) and the AdaBoost algorithm. Two fundamental ideas behind 
these algorithms are especially far-reaching. The first one is that we can 
transform many classical linear algorithms into highly flexible nonlinear 
algorithms by using kernel functions. The second one is that we can make 
accurate predictions by building an ensemble of models without much 
fine-tuning for each, rather than carefully fine-tuning a single model.

In this expository article, I first present the main ideas behind kernel 
methods (Section \ref{sec:kerns}) and ensemble methods (Section 
\ref{sec:ens}) by reviewing four prototypical algorithms: the support 
vector machine \citep[SVM, e.g.,][]{svmbk}, kernel principal component 
analysis \citep[kPCA,][]{kPCA}, AdaBoost \citep{boosting-orig}, and random 
forest \citep{randomForest}. I then illustrate the influence of these 
ideas on my own research (Section \ref{sec:mywork}) by highlighting two 
recent algorithms that I have invented with my collaborators: LAGO 
\citep{lago}, a fast kernel machine for rare target detection; and 
Darwinian evolution in parallel universes \citep{pga}, an ensemble method 
for variable selection.

To better focus on the main ideas and not be distracted by the 
technicalities, I shall limit myself mostly to the two-class 
classification problem, although the SVM, AdaBoost and random forest can 
all deal with multi-class classification and regression problems as well. 
Technical details that do not affect the understanding of the main ideas 
are also omitted.

\section{Kernels}
\label{sec:kerns}

I begin with kernel methods. Even though the idea of kernels is fairly 
old, it is the support vector machine (SVM) that ignited a new wave of 
research in this area over the past 10 to 15 years.

\subsection{SVM}
\label{sec:svm}

In a two-class classification problem, we have predictor vectors 
$\myvec{x}_i \in \mathbb{R}^d$ and class labels $y_i \in \{-1, +1\}$, $i = 
1, 2, ..., n$. SVM seeks an optimal hyperplane to separate the two 
classes.

A hyperplane in $\mathbb{R}^d$ consists of all $\myvec{x} \in 
\mathbb{R}^d$ that satisfy the linear equation:
\[
 f(\myvec{x}) = \gvec{\beta}^T  \myvec{x} + \beta_0 = 0.
\]   
Given $\myvec{x}_i \in \mathbb{R}^d$ and $y_i \in \{-1, +1\}$, a 
hyperplane is called
a separating hyperplane if there exists $c > 0$ such that
\beqn
\label{eq:separate}
 y_i (\gvec{\beta}^T  \myvec{x}_i + \beta_0 ) \geq c
 \quad
 \forall~i=1,2,...,n.
\eeqn
Clearly,
a hyperplane can be reparameterized by scaling, e.g.,
\[
 \gvec{\beta}^T  \myvec{x} + \beta_0 = 0
 \quad
 \mbox{is equivalent to}
 \quad
 s (\gvec{\beta}^T  \myvec{x} + \beta_0) = 0
\]
for any scalar $s$.
In particular, we can scale the hyperplane so that (\ref{eq:separate})
becomes
\beqn
\label{eq:separation}
 y_i(\gvec{\beta}^T \myvec{x}_i + \beta_0) \geq 1 \quad 
\forall~i=1,2,...,n,
\eeqn
that is, scaled so that $c=1$.
A separating hyperplane satisfying condition (\ref{eq:separation})
is called a {\em canonical} separating hyperplane (CSHP).

\begin{figure}[hptb]
 \centering
 \includegraphics[height=3in, angle=270]{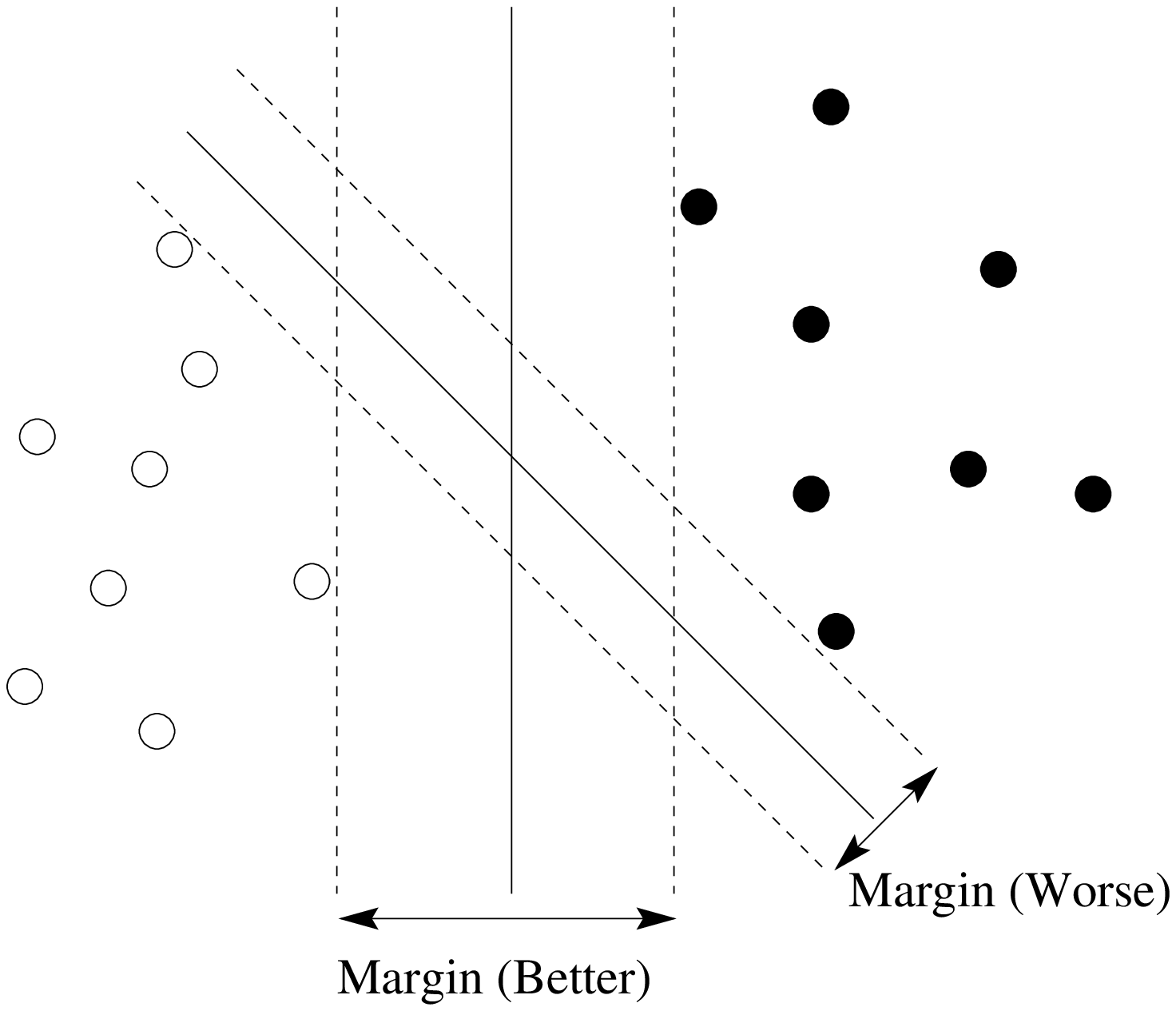}
 \mycap{Two separating hyperplanes, one with a larger
 margin than the other.}
\label{fig:margin}
\end{figure}

If two classes are perfectly separable, then there exist an infinite 
number of separating hyperplanes. Figure \ref{fig:margin} shows two 
competing hyperplanes in such a situation. The SVM is based on the notion 
that the ``best'' canonical separating hyperplane to separate two classes 
is the one that is the farthest away from the training points. This notion 
is formalized mathematically by the {\em margin} of a hyperplane --- 
hyperplanes with larger margins are better. In particular, the margin of a 
hyperplane is equal to
\[
 \mbox{margin} = 2 \times \min \{y_i d_i, i=1,2,...,n\},
\] 
where $d_i$ is the signed distance between observation $\myvec{x}_i$ and 
the hyperplane; see Figure \ref{fig:margin} for an illustration. Figure 
\ref{fig:margin} also shows to a certain extent why large margins are good 
on an intuitive level; there is also an elaborate set of 
theories to justify this \citep[see, e.g.,][]{vapnik}.

It can be shown
\citep[e.g.,][Section 4.5]{fht}
%[\ref{sec:details}\ref{ex:margin}]
that $d_i$ is equal to
\beqn
\label{eq:signdist}
 d_i = \frac{1}{\|\gvec{\beta}\|} (\gvec{\beta}^T \myvec{x}_i + \beta_0).
\eeqn
Then, equations (\ref{eq:separation}) and (\ref{eq:signdist}) together 
imply that the margin of a CSHP is equal to
\[
 \mbox{margin} = 2 \times \min\{y_i d_i\} = \frac{2}{\|\gvec{\beta}\|}.
\]
To find the ``best'' CSHP with the largest margin, we are 
interested 
in solving
the following optimization problem: 
\beqn
\label{eq:svmobj}
 \min \quad
 \frac{1}{2} \|\gvec{\beta}\|^2 + \gamma \dsum_{i=1}^n \xi_i
\eeqn
\beqn
\label{eq:svmconstraint}
 \mbox{subject to} \quad 
 y_i(\gvec{\beta}^T \myvec{x}_i + \beta_0) \geq 1 - \xi_i
 \quad \mbox{and} \quad
  \xi_i \geq 0 
 \quad \forall~i.  
\eeqn
The extra variables $\xi_i$ are introduced to
relax the separability condition (\ref{eq:separation}) 
because, in general, we can't assume the two classes are always perfectly 
separable.
The term $\gamma \sum \xi_i$ acts as a penalty to control the degree of
such relaxation, and $\gamma$ is a tuning parameter.

The main message from the brief introduction above is this: SVM tries to 
find the best CSHP; it is therefore a linear 
classifier. The usual immediate response to this message is: So what? How 
does this make the SVM much different from and superior to classical 
logistic regression?

Equivalently, the constrained optimization problem above can be written as 
\citep[e.g.,][Exercise 12.1]{fht}
%[\ref{sec:details}\ref{ex:svmlossequiv}]
\beqn
\label{eq:svmobj-hinge}
 \min \quad \sum_{i=1}^n
 \left[ 1 - y_i(\gvec{\beta}^T \myvec{x}_i + \beta_0) \right]_{+} 
 + \lambda \|\gvec{\beta}\|^2,
\eeqn
where 
\beqnn
[z]_{+} = \begin{cases}
 z & \mbox{if } z > 0, \\
 0 & \mbox{if } z \leq 0.
\end{cases}
\eeqnn
For statisticians, the objective function in (\ref{eq:svmobj-hinge}) has 
the familiar form of a loss function plus a penalty term.
For the SVM, the loss function is $[1-y(\gvec{\beta}^T 
\myvec{x}+\beta_0)]_{+}$, and it is indeed very similar to the 
binomial log-likelihood used by logistic regression 
\citep[e.g.,][Figure 12.4]{fht}.
%[\ref{sec:details}\ref{ex:svm-hinge}]. 
But the usual logistic regression model does not include the penalty term 
$\lambda \|\gvec{\beta}\|^2$. This is the familiar ridge penalty and often 
stabilizes the solution, especially in high-dimensional problems. Indeed, 
this gives the SVM an advantage.

However, one can't possibly expect a linear classifier to succeed in 
general situations, no matter how optimal the hyperplane is. So, why is 
the SVM such a sensational success? 

\subsection{The ``kernel trick''}

\citet[][Chapters 5 and 6]{svmbk} provided detailed derivations to show 
that the optimal $\gvec{\beta}$ looks like this: 
%[\ref{sec:details}\ref{ex:kkt}]:
\[
 \gvec{\beta} = \dsum_{i \in SV} \alpha_i y_i \myvec{x}_i,
\]
where ``SV'' denotes the
set of ``support vectors'' with $\alpha_i > 0$ strictly positive;
the coefficients $\alpha_i, i= 1,2,...,n$, are solutions to
the (dual) problem:
\beqn
\label{eq:svmdual}
 \max \quad \sum_{i=1}^n \alpha_i - \frac{1}{2}
 \sum_{i=1}^n\sum_{j=1}^n \alpha_i \alpha_j y_i y_j 
 \myvec{x}^T_i \myvec{x}_j
\eeqn
\beqn
\label{eq:svmdualconstraint}
 \mbox{s.t.} \quad \sum_{i=1}^n \alpha_i y_i = 0
 \quad\mbox{and}\quad
 \alpha_i \geq 0 \quad \forall~i.
\eeqn
This means
the resulting hyperplane can be written as
\beqn
\label{eq:svmsolun}
 f(\myvec{x}) =
 \gvec{\beta}^T \myvec{x} + \beta_0 =
 \dsum_{i \in SV} \alpha_i y_i
 \myvec{x}_i^T \myvec{x} + \beta_0 = 0.
\eeqn

The key point here is the following: In order to obtain $\alpha_i$, one 
solves (\ref{eq:svmdual})-(\ref{eq:svmdualconstraint}), a problem that 
depends on the predictors $\myvec{x}_i$ only through their inner-products 
$\myvec{x}^T_i \myvec{x}_j$; once the $\alpha_i$'s are obtained, the 
ultimate decision function (\ref{eq:svmsolun}) is also just a function of 
inner-products in the predictor space.

Therefore, one can make SVM a lot more general simply by defining a 
``different kind of inner-product,'' say, $K_h(\myvec{u}; \myvec{v})$, in 
place of $\myvec{u}^T \myvec{v}$. The function $K_h(\myvec{u};\myvec{v})$ 
is a called a kernel function, where $h$ is a hyper-parameter, which is 
often determined empirically by cross-validation.
Then, (\ref{eq:svmdual}) becomes
\beqn
\label{eq:svmdual-K}
 \max \quad \sum_{i=1}^n \alpha_i - \frac{1}{2}
 \sum_{i=1}^n\sum_{j=1}^n \alpha_i \alpha_j y_i y_j 
 K_h(\myvec{x}_i; \myvec{x}_j)
\eeqn
and the decision function (\ref{eq:svmsolun}) becomes
\beqn
\label{eq:svm-soln}
 f(\myvec{x}) = \dsum_{i \in SV} \alpha_i y_i
 K_h(\myvec{x}; \myvec{x}_i) + \beta_0 = 0.
\eeqn
The boundary is linear in the space of $\phi(\myvec{x})$ where
$\phi(\cdot)$ is such that 
\[ 
K_h(\myvec{u}; \myvec{v}) =
\phi(\myvec{u})^T \phi(\myvec{v}),
\]
but generally it is nonlinear in the original predictor space (unless one 
picks a linear kernel function). Mercer's theorem \citep{mercer} 
guarantees the existence of such $\phi(\cdot)$ as long as $K_h$ is a 
non-negative definite kernel function. The beauty here is that we don't 
even need to define the mapping $\phi(\cdot)$ explicitly; all we have to 
do is to pick a kernel function $K_h(\myvec{u};\myvec{v})$. This makes the 
SVM very general.

\subsection{Kernelization of linear algorithms}
\label{sec:k-trick}

That we can apply a linear method in a different space is, of course, not 
a new idea to statisticians at all. For example, we all know how to fit a 
high-order polynomial using linear regression --- simply add the terms 
$x^2, x^3, ..., x^d$ to the regression equation!

The idea that we don't need to explicitly create these high-order terms is 
perhaps somewhat less familiar. Actually, it is not really a new idea, 
either; it is less familiar only in the sense that students usually don't 
learn about it in ``Regression Analysis 101.''

However, the SVM does deserve some credit in this regard. Even though 
the basic idea of kernels is fairly old, it is the SVM 
that has revived 
it and brought it back into the spotlight for applied statisticians. The 
basic idea is as follows.
 
A typical data matrix we encounter in statistics, $\myvec{X}$, is $n
\times d$, stacking $n$ observations $\myvec{x}_1, \myvec{x}_2, ...,
\myvec{x}_n \in \mathbb{R}^d$ as $d$-dimensional row vectors. That is,
\[
 \myvec{X} =
 \bvector
 \myvec{x}_1^T \\ 
 \myvec{x}_2^T \\
 \vdots \\  
 \myvec{x}_n^T
 \evector.
\]
It is easy to see that
\[
 \myvec{X} \myvec{X}^T =
 \bvector
 \myvec{x}_1^T \\
 \myvec{x}_2^T \\
 \vdots \\
 \myvec{x}_n^T
 \evector
 \left(
 \myvec{x}_1 \quad
 \myvec{x}_2 \quad
 \hdots \quad
 \myvec{x}_n
 \right)
=
 \bfourmatrix
 \myvec{x}_1^T \myvec{x}_1 & \myvec{x}_1^T \myvec{x}_2
 & \hdots & \myvec{x}_1^T \myvec{x}_n \\
 \myvec{x}_2^T \myvec{x}_1 & \myvec{x}_2^T \myvec{x}_2
 & \hdots & \myvec{x}_2^T \myvec{x}_n \\
 \vdots & \vdots & \ddots & \vdots \\
 \myvec{x}_n^T \myvec{x}_1 & \myvec{x}_n^T \myvec{x}_2
 & \hdots & \myvec{x}_n^T \myvec{x}_n \\
 \efourmatrix
\] 
is an $n \times n$ matrix of pairwise inner-products. Therefore,
if a linear algorithm can be shown to depend on the data matrix
$\myvec{X}$ only through 
\beqn
\label{eq:k-innprodmat}
\myvec{K} \equiv \myvec{X} \myvec{X}^T,
\eeqn 
then it can be easily 
``kernelized'' --- we simply replace
each inner-product entry of $\myvec{K}$ with $K_{ij} = K_h(\myvec{x}_i,
\myvec{x}_j)$, where $K_h(\cdot, \cdot)$ is a desired kernel function.

\subsection{Kernel PCA}

Kernel principal component analysis \citep[kPCA;][]{kPCA} is a successful 
example of ``kernelizing'' a well-known classic linear algorithm. 
To focus on the main idea, let us
assume that the data matrix $\myvec{X}$ is already centered
so that each column has mean zero. Let 
\beqn
\label{eq:sample-cov}
\myvec{S} = \myvec{X}^T \myvec{X}.
\eeqn
Then, the (ordered) eigenvectors of $\myvec{S}$, say 
$\myvec{u}_1, \myvec{u}_2, ..., \myvec{u}_d$, are the principal
components. Being eigenvectors, they satisfy the equations
\beqn
\label{eq:PCAeig} 
 \myvec{S} \myvec{u}_j = \lambda_j \myvec{u}_j,
 \quad j = 1, 2, ..., d.
\eeqn
Equations (\ref{eq:sample-cov}) and (\ref{eq:PCAeig}) together lead to
\beqn 
\label{eq:PCAeig2}
\myvec{X}^T \myvec{X} \myvec{u}_j = \lambda_j \myvec{u}_j,
 \quad j = 1, 2, ..., d.
\eeqn
This shows that $\myvec{u}_j$ can be represented in the form of
$\myvec{X}^T \gvec{\alpha}_j$ --- by letting $\gvec{\alpha}_j = \myvec{X}
\myvec{u}_j/\lambda_j$, to be specific. 
We will plug $\myvec{u}_j = \myvec{X}^T \gvec{\alpha}_j$ into 
(\ref{eq:PCAeig2}) and reparameterize the eigenvalue problem in terms of 
$\gvec{\alpha}_j$.

For $j = 1, 2, ..., d$, this leads to
\beqn
\label{eq:PCAeig3}
\myvec{X}^T \myvec{X} \myvec{X}^T \gvec{\alpha_j} = \lambda_j
\myvec{X}^T \gvec{\alpha}_j.
\eeqn
If we left-multiply both sides by $\myvec{X}$, we get
\beqnn
\myvec{X} \myvec{X}^T \myvec{X} \myvec{X}^T \gvec{\alpha_j} = \lambda_j
\myvec{X} \myvec{X}^T \gvec{\alpha}_j,
\eeqnn  
or simply
\beqn
\label{eq:PCAeig4}
\myvec{K}^2 \gvec{\alpha_j} = \lambda_j \myvec{K} \gvec{\alpha}_j,
\eeqn
which shows that $\gvec{\alpha}_j$ can be obtained by solving a
problem 
that depends on the data matrix only through 
the inner-product matrix $\myvec{K}$.

\citet{kPCA} explained why, in the context of kPCA, it is sufficient to 
reduce (\ref{eq:PCAeig4}) to $\myvec{K} \gvec{\alpha}_j = \lambda_j 
\gvec{\alpha}_j$; I 
do not go into this detail here.
Once we obtain the $\gvec{\alpha}_j$'s, 
%[\ref{sec:details}\ref{ex:kPCAeig}],
suppose we'd like to project new data
$\myvec{X}_{new}$
onto a few leading principal components, e.g., 
$\myvec{X}_{new} 
\myvec{u}_j$. We immediately find that
\[
\myvec{X}_{new} 
\myvec{u}_j =
 \myvec{X}_{new} \myvec{X}^T \gvec{\alpha}_j,
\]
and it is easily seen that $\myvec{X}_{new} \myvec{X}^T$ is just a
matrix of pairwise inner products between each new and old observations.

\begin{figure}[hptb]
 \centering
 \includegraphics[height=2.5in]{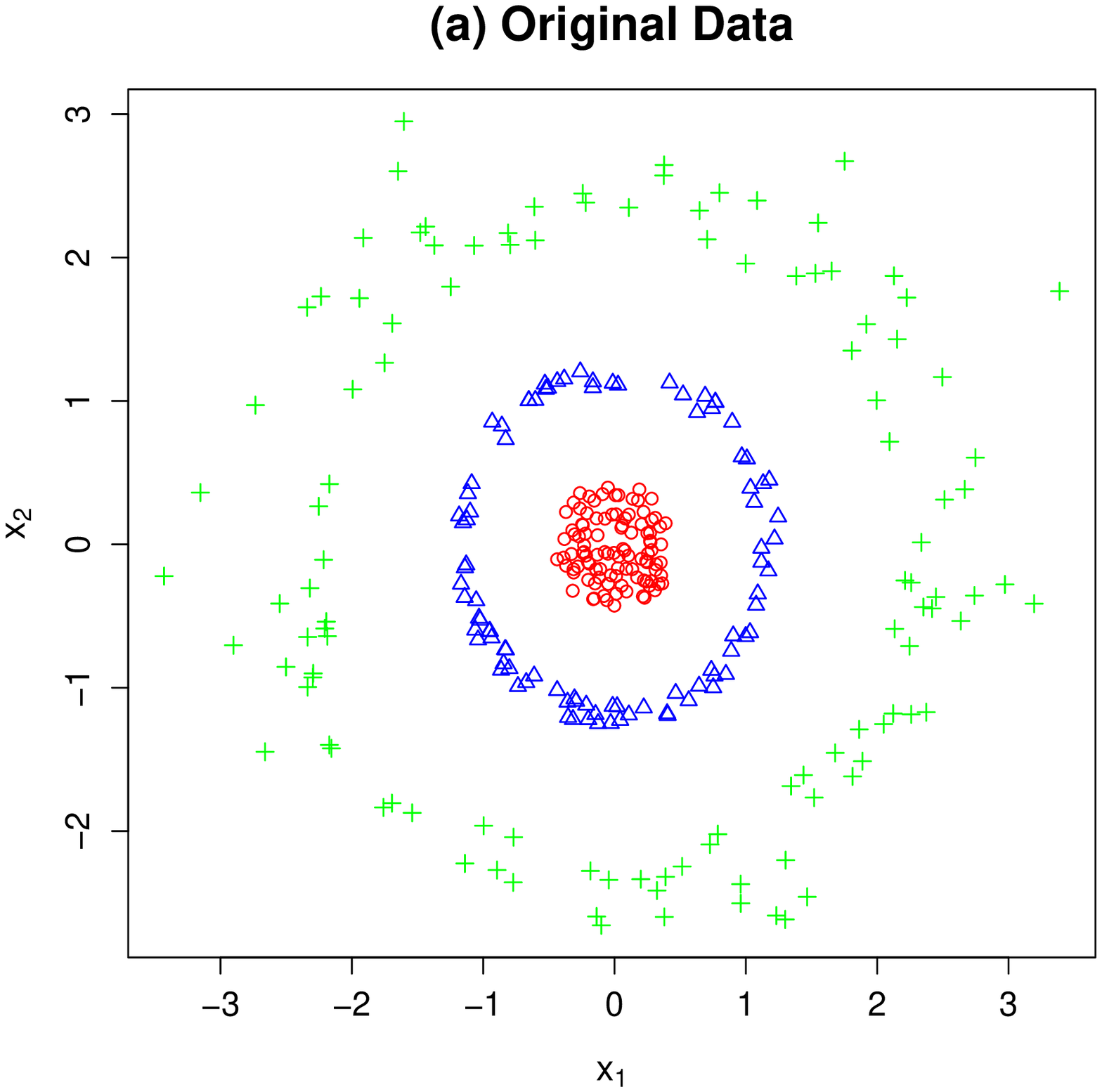}
 \includegraphics[height=2.5in]{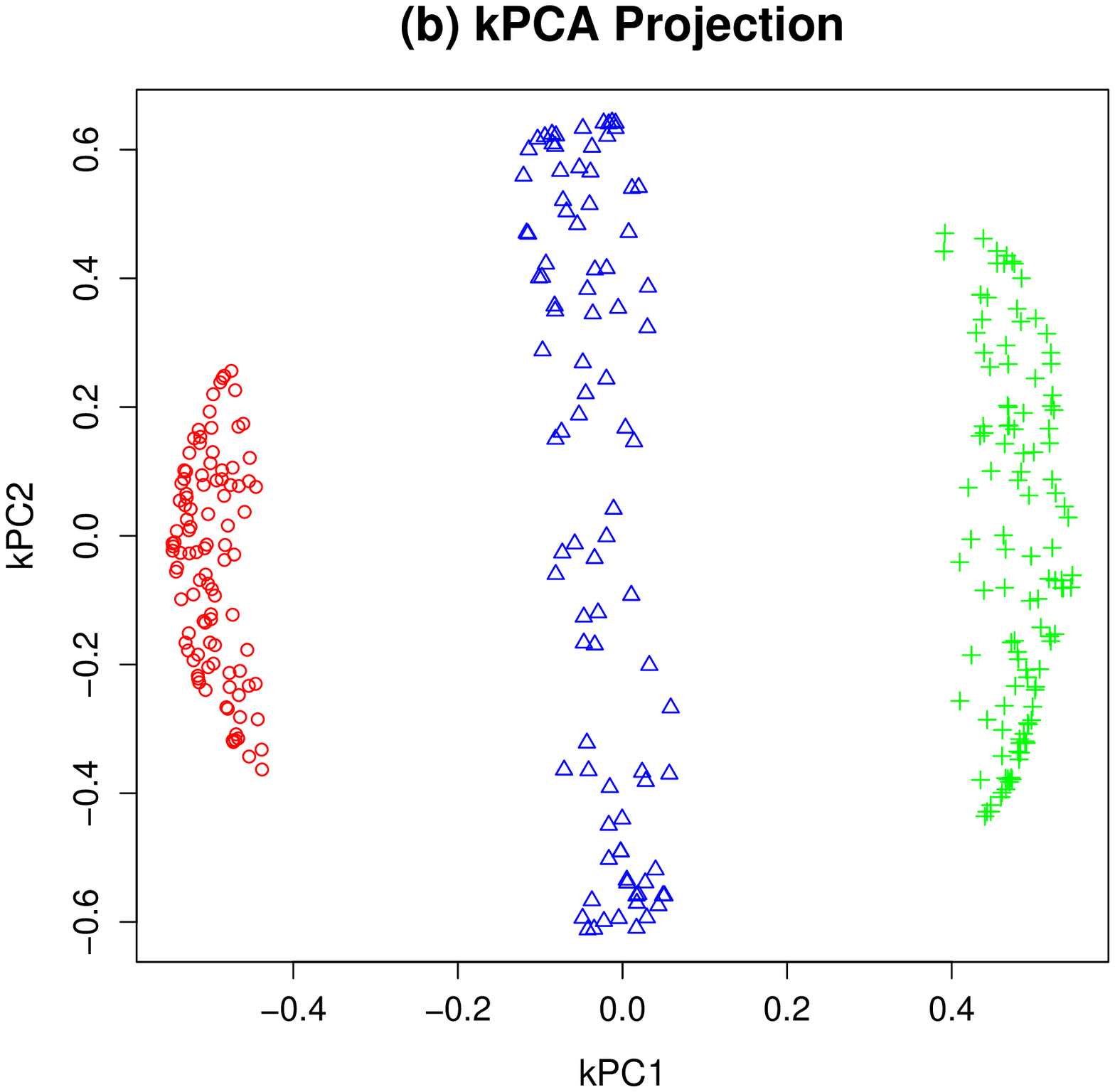}
  \mycap{Kernel PCA, toy example. (a) Original data. (b) 
  Projection onto the first two
  kernel principal components.} 
\label{fig:kPCA}
\end{figure}

Hence, it becomes clear that both finding and projecting onto principal 
components depend on just the inner-products and, according to 
Section~\ref{sec:k-trick}, PCA can be ``kernelized'' easily.
Figure \ref{fig:kPCA} shows a toy example. There are some spherical data 
in $\mathbb{R}^2$. The data being spherical, all directions have equal 
variance and there are no meaningful principal components in the 
traditional sense. But by using a 
Gaussian kernel --- equation (\ref{eq:RBFkerSVM}) below with $h=1$ --- in 
place of all the inner-products, the first kernel 
principal direction obtained gives a meaningful order of how far each 
observation is away from the origin. In this case, kernel PCA has 
successfully discovered the (only) underlying pattern in the data, one 
that is impossible to detect with classical PCA.
 
\subsection{Discussion: Kernel methods are like professional cameras}
\label{sec:kernsdisc}

Any acute reader must have noticed that, so far, I have never really 
discussed the kernel function $K_h(\myvec{u};\myvec{v})$ explicitly. This 
is not an accident. It is often claimed that one important advantage of 
these kernel methods lies in their modularity: to solve a different 
problem, just use a different kernel function. Any discussion about kernel 
functions, therefore, is best carried out in the context of a specific 
problem.

Of course, to be effective in practice, we must use the right kernel 
function. What's more, we must choose the right hyper-parameter $h$ as 
well, and the performance of the method can be quite sensitive to these 
choices in practice. These are no trivial tasks and often require a 
considerable amount of data analytic experience as well as knowledge of 
the specific application area.

In this regard, these kernel-based algorithms are very much like 
professional cameras. They are capable of producing great pictures even 
under very difficult conditions, but you need to give them to a 
professional photographer. If you give them to an amateur or novice, you 
can't expect great pictures. The photographer must know how to select the 
right lens, set the right shutter speed, and use the right aperture for 
any given condition. If any of these parameters is not set appropriately, 
the result could be a disaster. But that does not mean the camera itself 
is a poor piece of equipment; it simply means one must be adequately 
trained to operate it. Much of the power of these professional cameras 
lies precisely in the fact that they allow a knowledgeable and 
experienced user to control exactly how each single picture should be 
taken.

\subsubsection{Example: Spam data}
\label{sec:SVMspam}

As a very simple illustration, let us try to see how well the SVM can 
predict on the spam data set, available at 
\url{http://www-stat.stanford.edu/~tibs/ElemStatLearn/index.html}. There 
are a total of $n=4,601$ observations, each with a binary response and 
$d=57$ predictors. For more details about this data set, refer to the 
aforementioned web site. I use an \verb!R! package called \verb!e1071! to 
fit SVMs and use the kernel function 
\beqn
\label{eq:RBFkerSVM}
K_h(\myvec{u}; \myvec{v}) = 
\mbox{exp} \left\{- h\|\myvec{u} - \myvec{v}\|^2 \right\}.
\eeqn

\begin{figure}[hptb]
 \centering
 \includegraphics[height=2.75in]{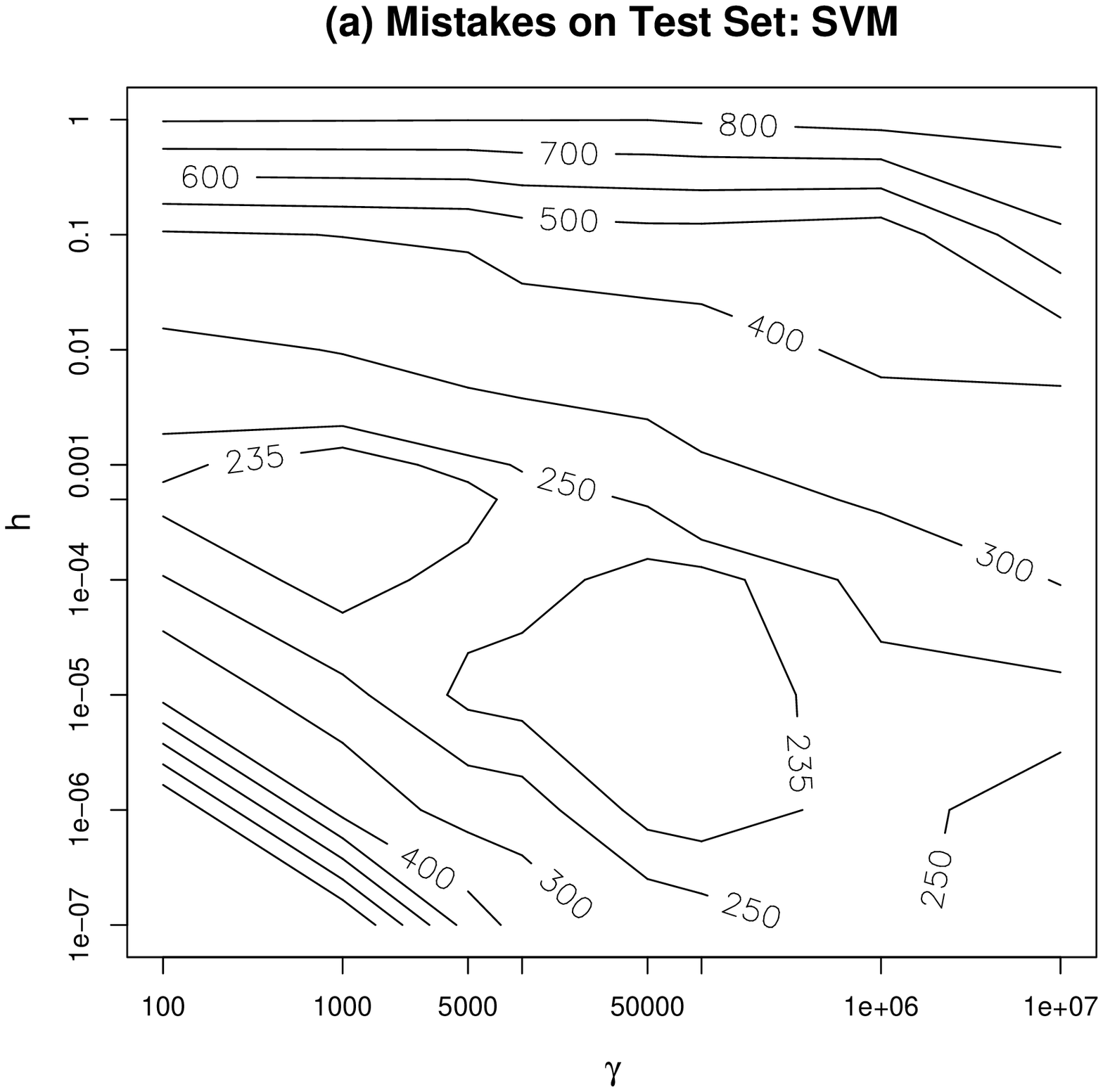}
 \includegraphics[height=2.75in]{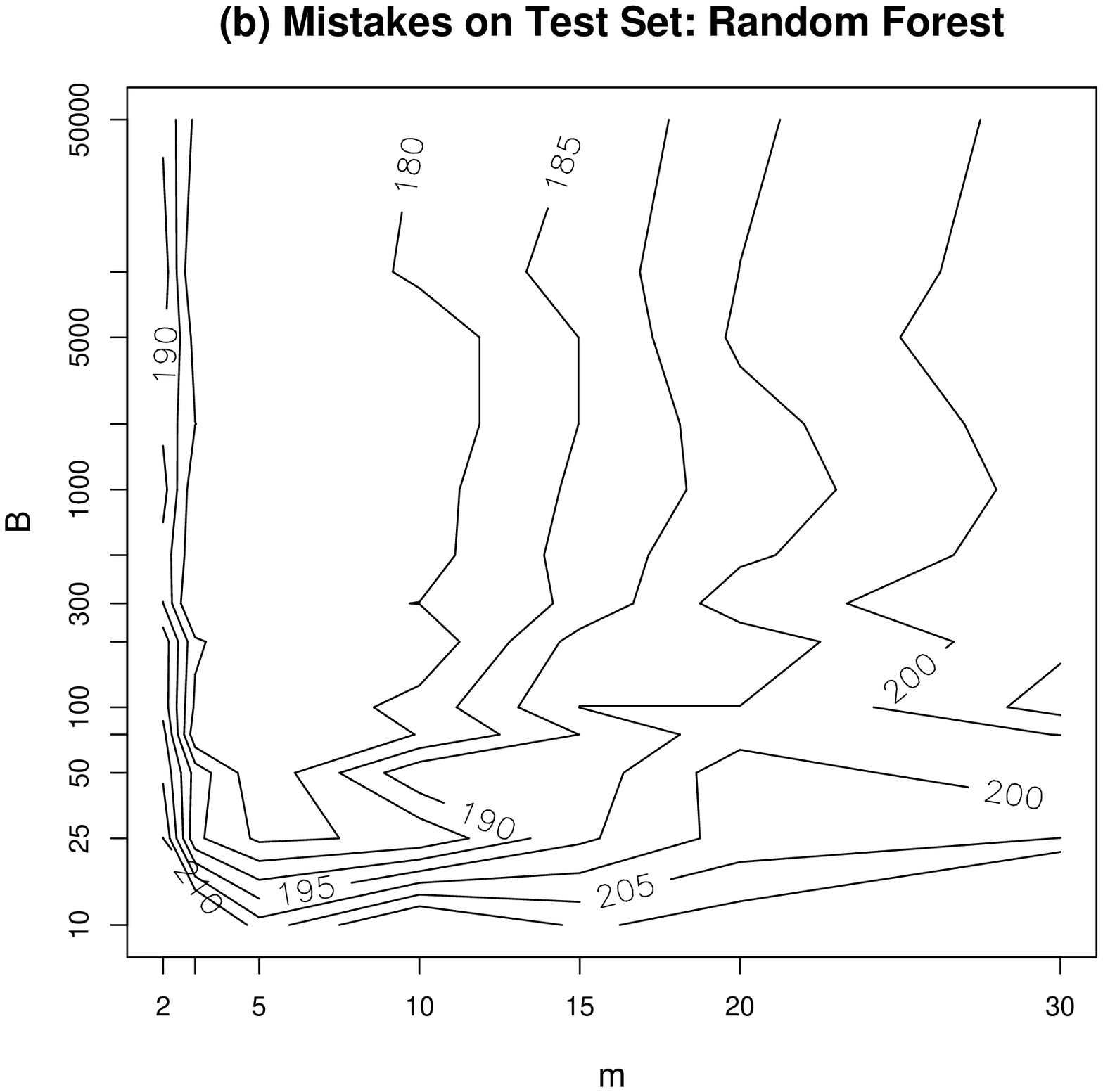}
 \mycap{Spam data example. (a) SVM: Number of misclassification errors on 
test 
data as a function of two tuning parameters, $\gamma$ and $h$ (see Section 
\ref{sec:SVMspam}). 
(b) Random forest: Number of misclassification errors on test data as a 
function of two tuning parameters, $m$ and $B$ (see Section 
\ref{sec:RFspam}).} 
\label{fig:spam} 
\end{figure}

A random sample of $1,536$ observations are used as training data and the 
remaining $3,065$ observations are used as test data. Using different 
values of $\gamma$ and $h$, a series of SVMs are fitted on the training 
data and then applied to the test data. The total number of 
misclassification errors on the test data are recorded and plotted for 
each pair of $(\gamma, h)$; see Figure \ref{fig:spam}(a). Here, $\gamma$ 
is the penalty parameter in equation (\ref{eq:svmobj}). 

Figure \ref{fig:spam}(a) shows that the performance of SVM using this 
particular
kernel function is very sensitive to the parameter $h$ but not as 
sensitive to the parameter $\gamma$. Given $h$, the prediction performance 
of SVM is often quite stable for a wide range of $\gamma$'s, but bad 
choices of $h$ can lead to {\em serious} deteriorations in the prediction 
performance. Therefore, if one uses the SVM without carefully tuning 
the parameter $h$, the result can be disastrous.

\section{Ensembles}
\label{sec:ens}

I now turn to ensemble methods. Again, I shall mainly focus on the 
two-class classification problem with predictor vectors $\myvec{x}_i \in 
\mathbb{R}^d$ and class labels $y_i \in \{-1, +1\}$, $i = 1, 2, ..., n$.

\subsection{AdaBoost}

AdaBoost constructs a collection of classifiers rather than one single 
classifier. The entire collection makes up an ensemble, and it is the 
ensemble --- not any single classifier alone --- that makes the final 
classification.

Table~\ref{tab:AdaBoost} contains an exact description of the AdaBoost 
algorithm. Here is a description of the algorithm in plain English:
Start by assigning equal weights to all observations in the training data. 
Sequentially build a series of classifiers. At each step, fit a 
classifier, say $f_b$, to the training data using the current weights. 
Calculate the (properly weighted) right-to-wrong ratio of this classifier; 
call it $R_b$. For those observations incorrectly classified by $f_b$, 
inflate their weights by a factor of $R_b$. With the new weights, build 
the next classifier. In the end, each classifier $f_b$ in the ensemble 
will cast a vote; its vote is to be weighted by the logarithm of its 
right-to-wrong ratio, $\log(R_b)$.

For people hearing about this algorithm for the very first time, AdaBoost 
certainly has a very strong mystical flavor to it. Intuitively, we can 
perhaps appreciate to some extent that the right-to-wrong ratio must be 
important for any classifier, but it is not at all clear why we should 
reweight incorrectly classified observations by this ratio each time, nor 
is it immediately clear why the final vote from each individual 
member of the ensemble should be weighted by the logarithm of this ratio.

\begin{table}[htpb]
\mycap{\label{tab:AdaBoost}%
The AdaBoost Algorithm.}
\centering
\fbox{%
\begin{tabular}{p{5in}}
\benum
\item Initial weights: $w_i = 1/n, \forall~i$. 

\item For $b = 1$ to $B$:

 \benum
 
\item Using weights $w_i, i=1,2,...,n$, fit a classifier 
  $f_b(\myvec{x}) \in \{-1, +1\}$. 

 \item
 \label{BoostStep:calc}
 Set
 \[
 \epsilon_b = \frac{\sum_{i=1}^n w_i I(y_i \neq f_b(\myvec{x}_i))}
  {\sum_{i=1}^n w_i},
 \quad
 R_b = \frac{1-\epsilon_b}{\epsilon_b},
 \quad
 a_b = \log(R_b).
 \] 

 \item\label{BoostStep:update}
 Update weights: $w_i \leftarrow w_i \times R_b$ 
 if $y_i \neq 
  f_b(\myvec{x}_i)$. 

 \eenum

\item[] End For.

\item Output an ensemble classifier
\[
 F(\myvec{x}) = 
 \mbox{sign}\left( \sum_{b=1}^B a_b f_b(\myvec{x}) \right).
\] 
\eenum
\end{tabular}}
\end{table}

This is no easy mystery to untangle. \citet{boosting} gave a very nice 
argument and revealed that the AdaBoost algorithm actually minimizes an 
exponential loss function 
using a forward stagewise approach. In 
particular, AdaBoost 
chooses the best $a_b$ and $f_b$ one
step at a time to minimize 
\[
 \sum_{i=1}^n \mbox{exp}
 \left( -y_i \sum_{b=1}^B a_b f_b(\myvec{x}_i)
 \right),
\]
which they showed to be very similar to maximizing the binomial 
log-likelihood.
%[\ref{sec:details}\ref{ex:adaboost}]. 
This particular interpretation has 
not only untangled the AdaBoost mystery (at least to some extent), but 
also led to many new (and sometimes better) versions of boosting 
algorithms.

\subsection{Random forest}

Professor Leo Breiman came up with the 
same basic idea of using a collection or an ensemble of models to make 
predictions, except he constructed his ensemble in a slightly different 
manner. Breiman called his ensembles {\em random forests}; details are 
given in Table~\ref{tab:RF}.

\begin{table}[htpb]
\mycap{\label{tab:RF}%
Breiman's Random Forest Algorithm.}
\centering
\fbox{%
\begin{tabular}{p{5in}}
\benum
\item For each $b = 1$ to $B$, fit a maximal-depth tree, $f_b(\myvec{x})$, as 
follows:

 \benum

 \item\label{RFstep:boot}
(Bootstrap Step) 
Draw a bootstrap sample of the training data; call it $D^{*b}$.
Use $D^{*b}$ to 
fit $f_b$.

 \item\label{RFstep:subset}
 (Random Subset Step) 
When building $f_b$,
randomly select a subset of $m<d$ predictors before making each split --- 
call it $S$, and make the best split over the set $S$ rather than 
over all possible predictors.

 \eenum

\item[] End For.

\item Output an ensemble classifier, i.e., to classify $\myvec{x}_{new}$, 
simply take majority vote over $\{f_b(\myvec{x}_{new}), b = 1, 2, ..., 
B\}$.
\eenum
\end{tabular}}
\end{table}

The history behind Breiman's random forest is very interesting. In 1996, 
he first proposed an ensemble algorithm called Bagging \citep{bagging}, 
which is essentially the random forest algorithm with just the bootstrap 
step (Table \ref{tab:RF}, step \ref{RFstep:boot}). In 2001, he added the 
random subset step (Table \ref{tab:RF}, step \ref{RFstep:subset}) and 
created random forest \citep{randomForest}.

Why did he add the extra random subset step?

\subsection{Breiman's theorem}
\label{sec:BreimanThm}

\citet{randomForest} proved a remarkable theoretical result. First, he 
gave a formal definition of random forests: The set 
\[
 \{f(\myvec{x}; \theta_b): \theta_b \overset{iid}{\sim}
 \mathcal{P}_{\theta}, b = 1, 2, ... B\}
\]
is called a random forest.

This definition requires some explanation. Here, $f(\myvec{x}; \theta_b)$
is a classifier completely parameterized by $\theta_b$. For example, if  
$f(\cdot; \theta_b)$ is a classification tree, then the parameter
$\theta_b$ specifies all the splits and the estimates in the terminal
nodes.
Next, the statement ``$\theta_b \overset{iid}{\sim}
\mathcal{P}_{\theta}$'' means that each $f(\cdot;\theta_b)$ is generated
independently and identically from some underlying random mechanism,
$\mathcal{P}_{\theta}$.

To be specific, in Breiman's implementation, iid sampling from the random 
mechanism $\mathcal{P}_{\theta}$ consists of: (i) iid sampling from the 
empirical distribution $F_n$ (the bootstrap step), and (ii) iid sampling 
from the set $\{1, 2, ..., d\}$ (the random subset step).

Breiman then 
proved that the prediction error of a random forest, $\epsilon_{RF}$, 
satisfies the inequality
\beqn
\label{eq:RFtheorem}
\epsilon_{RF} \leq \bar{\rho} \left( \frac{1-s^2}{s^2} \right),
\eeqn
where $\bar{\rho}$ is the mean correlation between any two members of the 
forest (ensemble) and $s$, the mean strength of a typical member of the 
forest (ensemble). This result --- including the exact definitions of 
$\bar{\rho}$ and $s$ --- is fairly technical; details can be found in 
\citet{randomForest}.
%[\ref{sec:details}\ref{ex:RF}; Appendix \ref{sec:RFappdx}]. 
Moreover, the actual 
bound itself is often useless. 
For example, 
if $s =
0.4$ and $\bar{\rho} = 0.5$, then 
one gets
\[
 \epsilon_{RF} \leq \bar{\rho} \left( \frac{1-s^2}{s^2} \right) =   
 0.5 \left( \frac{1-0.4^2}{0.4^2} \right) = 2.625,
\]
but of course the error rate is less than 100\%.

So, why is this result significant?

\subsection{The secret of ensembles}
\label{sec:secret}

The fundamental idea of using an ensemble classifier rather than a single 
classifier is nothing short of being revolutionary. It also is remarkable 
that building these ensembles is often {\em relatively} mindless. Take 
Breiman's random forest, for example. There is no need to prune the 
individual trees.

Clearly, there are many different ways to build an ensemble, AdaBoost and 
Breiman's random forest being two primary examples. What's the most 
effective way?

Recall the formal definition of random forests. The random mechanism 
$\mathcal{P}_{\theta}$ that generates the individual members of the forest 
is unspecified. You are free to pick any mechanism you want. Surely some 
mechanisms are bound to be more effective than others. What's the most 
effective mechanism?

Breiman's result is significant because it tells us what makes a good 
random forest. Breiman's theorem (\ref{eq:RFtheorem}) tells us that a good 
random forest should have a small $\bar{\rho}$ and a large $s$. That is, 
we should try to reduce the correlation between individual classifiers 
within the ensemble and make each individual classifier as accurate as 
possible.

This explains why Breiman added the random subset step into his original 
Bagging algorithm: extra randomness is needed to reduce the correlation 
between individual trees; the bootstrap step alone is not enough!

Interestingly, we can see that AdaBoost actually
operates in a similar way. 
Going back to step (\ref{BoostStep:calc}) in Table \ref{tab:AdaBoost}, we 
have 
\[
 \epsilon_b = \frac{\sum_{i=1}^n w_i I(y_i \neq f_b(\myvec{x}_i))}
  {\sum_{i=1}^n w_i}.
\]
From this, we can write
\[
  \epsilon_b \sum_{all} w_i = \sum_{wrong} w_i
  \quad\mbox{and}\quad
  (1-\epsilon_b) \sum_{all} w_i = \sum_{right} w_i,
\]
where ``all'' means $i=1, 2, ..., n$; ``wrong'' denotes the set $\{i: y_i 
\neq f_b(\myvec{x}_i)\}$
and ``right,'' the set $\{i: y_i = f_b(\myvec{x}_i)\}$.
Step (\ref{BoostStep:update}) in Table \ref{tab:AdaBoost} gives the 
explicit update rule;
the new weights are:
\beqnn
 w_i^{new} = 
 \begin{cases}
 w_i \times
 \left(\frac{\displaystyle 1-\epsilon_b}{\displaystyle \epsilon_b} 
 \right), & \mbox{ for } i \in \mbox{wrong}; \\
 w_i, & \mbox{ for } i \in \mbox{right}.
 \end{cases}
\eeqnn
Therefore, we can see that
\begin{multline*}
 \sum_{wrong} w_i^{new} = 
 \left(\frac{1-\epsilon_b}{\epsilon_b} 
 \right) \sum_{wrong} w_i 
 = 
 (1-\epsilon_b)\sum_{all} w_i
 = \sum_{right} w_i
 = \sum_{right} w_i^{new},
\end{multline*}
which means the misclassification error of $f_b$ under the new weights 
$w_i^{new}$ is 
exactly 50\% --- the worst possible error. 

The next classifier, $f_{b+1}$, is built using these new weights, so it is 
set up to work with a (weighted) dataset that the current classifier, 
$f_b$, cannot classify. This is sometimes referred to as ``decoupling'' in 
the boosting literature --- the classifier $f_{b+1}$ is decoupled from 
$f_b$.

In Breiman's language, we can say that the adaptive and hitherto 
mysterious reweighting mechanism in AdaBoost is actually aiming to reduce 
the correlation between consecutive members of the ensemble.

\subsection{Discussion: Ensemble methods are like foolproof cameras}
\label{sec:ensdisc}

Compared with kernel methods, ensemble methods are very much like 
foolproof cameras. They are relatively easy for the less experienced users 
to operate. This does not mean they don't have any tuning parameters; they 
do. Even when using a foolproof camera, one must still make a few 
decisions, e.g., whether or not to turn on the flash, and so on. But 
relatively speaking, the number of decisions one has to make is much more 
limited and these decisions are also relatively easy to make.

For example, in Breiman's random forest, the size of the subset, $m$ 
(Table~\ref{tab:RF}, step~\ref{RFstep:subset}), is an important tuning 
parameter. If $m$ is too large, it will cause $\bar{\rho}$ to be too 
large. In the extreme case of $m=d$, all the trees in the forest will be 
searching over the entire set of variables in order to make splits, and 
they will be identical --- since the tree-growing algorithm is 
deterministic conditional on the data. On the other hand, if $m$ is too 
small, it will cause $s$ to be too small. In the extreme case of $m=1$, 
all the trees will essentially be making random splits, and they will not 
be very good classifiers. There is plenty of empirical evidence to 
suggest, however, that the parameter $m$ is still relatively easy to 
choose in practice. Moreover, the parameter $m$ is not as sensitive as the 
complexity parameter $h$ of a kernel function (also see Section 
\ref{sec:RFspam} below). Translation: Even if you are a bit off, the 
consequences will not be quite so disastrous.

I have had many occasions working with graduate students trying to make 
predictions using the SVM and Breiman's random forest. They {\em almost 
always} produce much better predictions with the random forest, even on 
problems that are well-suited for the SVM! Sometimes, their SVMs actually 
perform worse than linear logistic regression. Certainly, there are many 
cases in practice where one would not expect the SVM to be much superior 
to linear logistic regression, e.g., when the true decision boundary is in 
fact linear. But if used correctly, the SVM should at least be comparable 
with linear logistic regression; there is no reason why it ever would be 
much worse. These experiences remind me over and over again just how 
difficult it can be for a novice to use the SVM.

But, as I stated in Section \ref{sec:kernsdisc}, you can't blame the 
professional camera if you don't know how to use it properly. There is 
always a tradeoff. With limited flexibility, even a fully-experienced 
professional photographer won't be able to produce images of the highest 
professional quality with just a foolproof camera, especially under 
nonstandard and difficult conditions. That's why professional cameras are 
still on the market. But we have to admit: {\em most} consumers are 
amateur photographers and, more often than not, they are taking pictures 
under fairly standard conditions. That's why the demand for foolproof 
cameras far exceeds that for professional cameras. I think the 
demand for statistical tools follows a similar pattern.

\subsubsection{Example: Spam data (continued)}
\label{sec:RFspam}

As a simple illustration, let us take a look at how well the random forest 
can predict on the spam data set. I use exactly the same set-up as in 
Section \ref{sec:SVMspam} and the \verb!randomForest! package in \verb!R!. 
Using different values of $m$ and $B$, a series of random forests are 
fitted on the training data and then applied to the test data. The total 
number of misclassification errors on the test data are recorded and 
plotted; see Figure \ref{fig:spam}(b). Here, we can see that the 
performance of random forests is more sensitive to the parameter $m$ than 
to the parameter $B$. Given $m$, the prediction performance of random 
forests is fairly stable as long as $B$ is sufficiently large, e.g., 
$B>100$ in this case. But it is important to use an $m$ that is neither 
too small nor too big, e.g., $3<m<10$ in this case.

However, if we compare panels (a) and (b) in Figure~\ref{fig:spam}, we can 
see that choosing the right $h$ for SVM is much more critical than 
choosing the right $m$ for random forest; performance deterioration is 
much more serious for bad choices of $h$ than for bad choices of $m$.

It is also clear from Figure \ref{fig:spam} that, for this particular data 
set, an SVM with kernel function (\ref{eq:RBFkerSVM}) is not 
competitive against a random forest, even if well tuned. In order to be 
competitive, it is necessary to use a different kernel function. I 
do not pursue this possibility here because getting the SVM to work for 
this data set is far from the main point of our discussion, but this 
example does demonstrate that choosing the right kernel function $K_h$ and 
picking the right hyperparameter $h$ are very important, and that an 
ensemble method such as the random forest can be somewhat easier to use in 
this regard.

\section{Perspectives}
\label{sec:mywork}

I now share a few personal perspectives on statistical learning research. 
Here, I am working with a particular definition of the word 
``perspective'' from the American Heritage Dictionary: a {\em 
subjective} evaluation of relative significance [emphasis added].

\subsection{Statistical learning research}
\label{sec:myopinion}

My discussions in Sections \ref{sec:kernsdisc} and \ref{sec:ensdisc} have 
led me to ask the following question: If I were the president of a big 
camera manufacturing company, how would I run such a business? Other than 
standard business divisions such as accounting and human resources, 
I see three main lines of operation:
\benum

\item (Consulting and Consumer Outreach)
Advise and teach photographers how to use various products and how 
to use the right equipment to produce great pictures under various 
difficult conditions. This is my consulting and consumer outreach 
division.

\item (High-end R\&D)
Understand the need of professional photographers and manufacture new, 
specialized equipment still lacking on the market. This is my 
R\&D division for my high-end 
consumers.

\item (Mass R\&D)
Build the next-generation foolproof camera. This is my R\&D division
for my mass consumers. 
\eenum
I see a great deal of parallelism in statistical learning research.
For statistical learning research, the 
consulting and consumer outreach 
division applies different learning methods to solve various difficult 
real-world problems; the high-end R\&D division develops new, specialized 
algorithms for analyzing new types of data or data with special 
characteristics; and the mass R\&D division 
develops better off-the-shelf learning algorithms.

With this particular point of view in mind, I end this article by briefly 
describing two personal learning products: a new kernel method from my 
high-end R\&D division, and a new ensemble method from my mass R\&D 
division.

\subsection{A high-end R\&D product: LAGO}
\label{sec:lago}

Consider a two-class problem in which the class of interest ($C_1$) is 
very rare; most observations belong to a majority, background class 
($C_0$). Given a set of unlabelled observations, the goal is 
to rank those belonging to $C_1$ ahead of the rest. 

Of course, one can use any classifier to do this as long as the classifier 
is capable of producing not only a class label but also an estimated 
posterior 
probability $P(y \in C_1 | \myvec{x})$ or a classification score. For 
example, the SVM does not estimate posterior probabilities, but the final 
decision function (\ref{eq:svm-soln}) is a classification score which can 
be used (at least operationally) to rank unlabelled observations --- 
whether this is effective or not is a separate issue.

\subsubsection{RBFnets}

The final decision function produced by SVM (\ref{eq:svm-soln}) is of the 
form
\beqn
\label{eq:RBFnet}
f(\myvec{x}) = \beta_0 + \sum_{\gvec{\mu}_i \in S} \beta_i
\phi(\myvec{x}; \gvec{\mu}_i, \myvec{R}_i),
\eeqn
where $\phi(\myvec{x}; \gvec{\mu}, \myvec{R})$ is a kernel function. For 
example, we can take $\myvec{R}$ to be diagonal and let $\phi$ be the 
Gaussian 
kernel 
\begin{eqnarray}
\label{eq:gaussker}
 \phi(\myvec{x}; \gvec{\mu}, \myvec{R}) =
 \frac{1}{\sqrt{(2\pi)^d} |\myvec{R}|}
 \mbox{exp}\left[{-\frac{\displaystyle (\myvec{x}-\gvec{\mu})^T 
\myvec{R}^{-2}   
   (\myvec{x}-\gvec{\mu})}{\displaystyle 2}} \right],
\end{eqnarray}  
where $|\myvec{R}|$ is the determinant of $\myvec{R}$. 

The function (\ref{eq:RBFnet}) is sometimes called a (single-layer) radial 
basis function network (RBFnet).
Generally speaking, to construct an RBFnet one must compute and specify 
three ingredients:
\begin{itemize}
\item[$\gvec{\mu}_i$,] the location parameter of each kernel
function --- together, they make up the set $S$;
\item[$\myvec{R}_i$,] the shape parameter of each kernel 
function; and
\item[$\beta_i$,] the coefficient in front of each kernel
function.
\end{itemize}   
Typically, one first specifies $\gvec{\mu}_i$ and $\myvec{R}_i$ and then
estimates the $\beta_i$'s 
by least-squares or maximum likelihood. Often, one sets 
$\myvec{R}_i = r\myvec{I}$ and treats the parameter $r$ as a global tuning 
parameter --- this is what SVM does. Determining the 
$\gvec{\mu}_i$'s or the best set $S$ from training data, however, is an
NP-hard 
combinatorial optimization problem in general.

The SVM can be viewed as an algorithm for determining the set 
$S$ and the $\beta_i$'s simultaneously \citep{svmrbf}; the set $S=SV$ is 
simply the set of all support vectors. In order to do so, SVM solves a 
quadratic programming instead of a combinatorial optimization problem.
%[\ref{sec:details}\ref{ex:kkt}].

\subsubsection{LAGO}

The product from my R\&D division is an algorithm called LAGO 
\citep{lago}.
The decision function constructed by LAGO for ranking 
unlabelled observations is as follows:
\beqn
\label{eq:LAGO}
f(\myvec{x}) = \sum_{\myvec{x}_i \in C_1} |\myvec{R}_i| 
\phi(\myvec{x}; \myvec{x}_i, \alpha\myvec{R}_i).
\eeqn
The parameter $\alpha$ is a global tuning parameter. 
In the simplest case, we take
\beqn
\label{eq:LAGO-radius}
\myvec{R}_i = r_i \myvec{I},
\eeqn
where $r_i$ is the 
average distance between the kernel center, $\myvec{x}_i \in C_1$, and its
$K$-nearest neighbors from $C_0$, i.e.,
\beqn
\label{eq:distanceC1-C0}
 r_{i} = \frac{1}{K} \sum_{\myvec{w} \in N_0(\myvec{x}_i, K)} 
 d(\myvec{x}_{i}, \myvec{w}).
\eeqn
The notation ``$N_0(\myvec{x}_i, K)$'' denotes the $K$-nearest neighbors 
of $\myvec{x}_i$ from $C_0$; and $d(\myvec{u},\myvec{v})$ is a distance 
function, e.g., $d(\myvec{u},\myvec{v}) = \|\myvec{u}-\myvec{v}\|$. 

By comparing (\ref{eq:LAGO})-(\ref{eq:LAGO-radius}) with 
(\ref{eq:RBFnet}), we can easily see that LAGO can also be viewed as an 
algorithm for constructing an RBFnet, just like the SVM. In particular, 
the three ingredients of the RBFnet are specified as follows:

\begin{itemize}

\item[$\gvec{\mu}_i$:] Every $\gvec{\mu}_i$ is a training observation 
$\myvec{x}_i$ from the rare class, $C_1$.

\item[$\myvec{R}_i$:] Each kernel function $\phi$ is spherical 
with radius proportional to the average distance between its center 
$\gvec{\mu}_i \in C_1$ and its the $K$-nearest neighbors from $C_0$.

\item[$\beta_i$:] Simply set $\beta_0 = 0$ and $\beta_i = 
|\myvec{R}_i|~\forall~i>0$.

\end{itemize}
Here we see that the only computation needed is the calculation of $r_i$; 
all other ingredients are completely determined a priori. The calculation 
of $r_i$, of course, is considerably simpler than quadratic programming, 
making LAGO {\em many times} faster and simpler than the SVM. Instead of 
solving an optimization problem to find support vectors, LAGO fully 
exploits the special nature of these rare-class detection problems and 
simply uses all training observations from the rare class 
as its ``support vectors,'' a significant shortcut.
Our empirical experiences show that the shortcut is highly worthwhile. We 
find that LAGO almost always performs as well as and sometimes even better 
than the SVM for these rare-class classification and detection problems.

\citet{lago} give a few theoretical arguments for why all these shortcuts 
are justified. Suppose $p_1(\myvec{x})$ and $p_0(\myvec{x})$ are density 
functions of $C_1$ and $C_0$. The main argument is that (\ref{eq:LAGO}) 
can be viewed as a kernel density estimate of $p_1$ adjusted locally by a 
factor that is approximately inversely proportional to $p_0$, i.e., 
$|\myvec{R}_i|$. The resulting ranking function $f(\myvec{x})$ is thus 
approximately a monotonic transformation of the posterior probability that 
item $\myvec{x}$ belongs to the rare class.

\begin{figure}[hptb]
 \centering
 \includegraphics[height=2.5in]{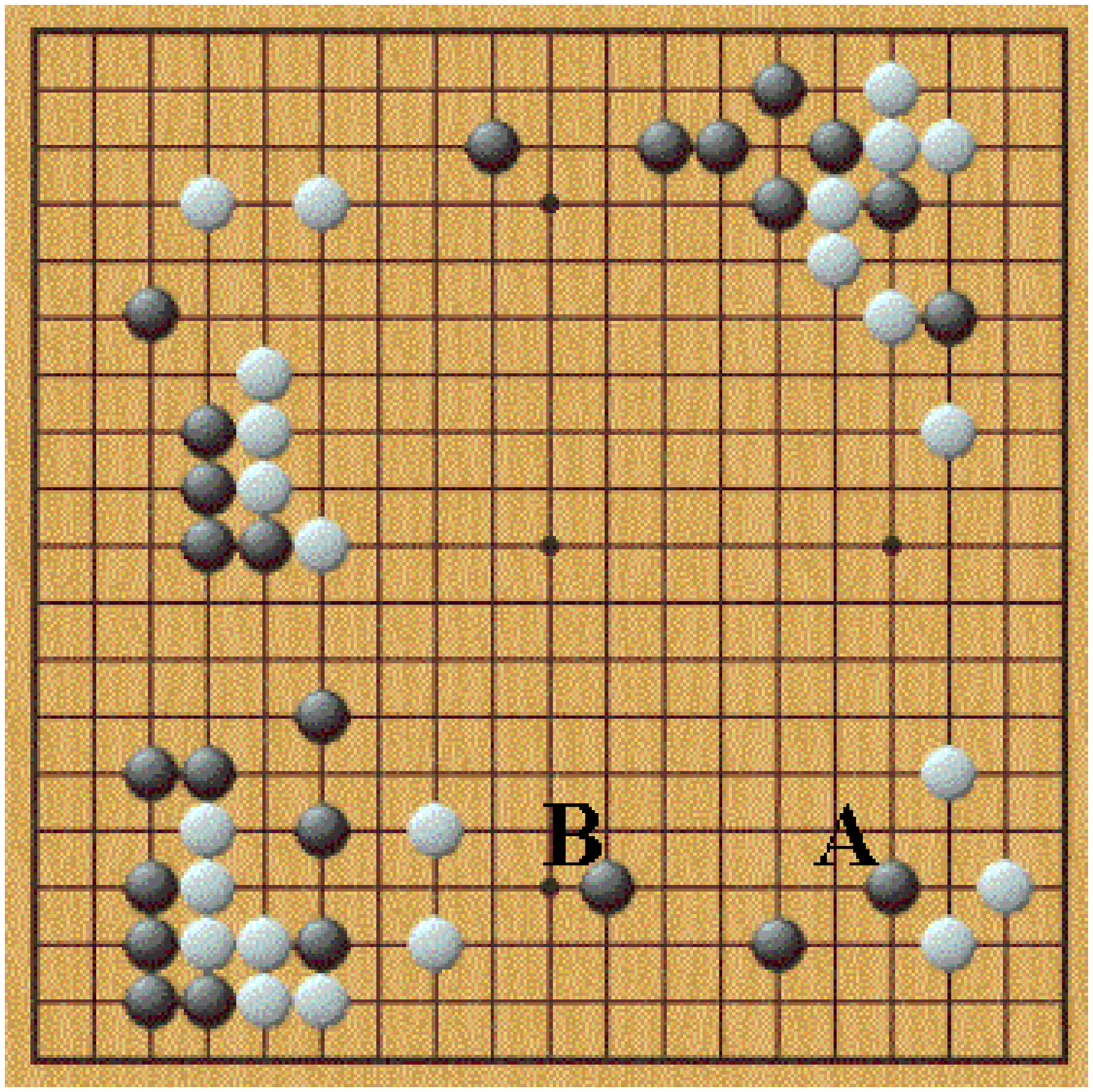}
 \mycap{The board game of GO. In this illustration, the black stone B 
 controls more territory than the black stone A.}
\label{fig:GO}
\end{figure}

The only nontrivial calculation performed by the algorithm, equation 
(\ref{eq:distanceC1-C0}), is somewhat special and nonstandard. The 
original idea came from a Chinese board game called GO.
Consider the two black stones labelled A and B in Figure~\ref{fig:GO}. A 
GO player will tell you that B controls more territories on the board 
than A. Why? Because, when compared with B, A is closer to more enemy 
(white) stones. Therefore, imagine two classes fighting for control over a 
common space. Given an observation from $C_1$, if we want to use a kernel 
function to describe its effective control over the entire space, we 
should use a large kernel radius if its nearby neighbors from $C_0$ are a 
long distance away and a small kernel radius if its nearby neighbors from 
$C_0$ are a short distance away. 
Equation (\ref{eq:distanceC1-C0}) captures
this basic principle.

\subsubsection{eLAGO versus sLAGO}

Instead of (\ref{eq:LAGO-radius})-(\ref{eq:distanceC1-C0}), the original 
LAGO paper \citep{lago} 
used
\beqn
\label{eq:eLAGO-radius}
\myvec{R}_i = \mbox{diag}\{r_{i1}, r_{i2}, ..., r_{id}\},
\quad
r_{ij} = \frac{1}{K} \sum_{\myvec{w} \in N_0(\myvec{x}_i, K)} |x_{ij} - 
w_{j}|.
\eeqn
That is, the kernel function $\phi$ was chosen to be elliptical rather 
than spherical. To distinguish the two, we call (\ref{eq:eLAGO-radius}) 
eLAGO and (\ref{eq:LAGO-radius}), sLAGO. For many real-world rare-class 
problems, the dataset often contains a limited amount of information 
because $C_1$ is very rare. As such, the extra flexibility afforded by 
eLAGO is seldom needed in practice. 

\subsubsection{Discussion: LAGO is a specialized kernel method}

LAGO is a kernel method, much like the SVM. There are two tuning 
parameters, $K$ and $\alpha$. Experiments similar to those described in 
Section \ref{sec:SVMspam} and Figure \ref{fig:spam} have shown that the 
performance of LAGO is not very sensitive to $K$ and much more sensitive 
to $\alpha$. In practice, it often suffices to fix $K=5$. 

LAGO is not a general-purpose method; it is a specialized algorithm for a 
special learning problem, namely rare-class classification and detection. 
Its main advantages are its speed and simplicity. Discussions in Section 
\ref{sec:kernsdisc} have made it clear that these kernel methods must be 
carefully tuned, e.g., using empirical procedures such as 
cross-validation. This means that, in practice, one almost always has to 
run these algorithms repeatedly for many times. One may be tempted to 
think that, if one algorithm takes 10 minutes to run and another takes 1 
minute, the difference is still ``negligible'' for all practical purposes, 
but such ten-fold differences are often significantly magnified if one has 
to run these two algorithms repeatedly for many times.

Apart from these practical matters such as time savings, the more 
important lesson from this research lies in the basic principles behind 
the construction of LAGO (\ref{eq:LAGO}). Here, we see that it always pays 
to exploit the special nature of an underlying problem. For these 
rare-class problems, there is only limited amount of useful information in 
the training data. LAGO fully exploits this fact by immediately zooming 
into the useful information (i.e., $\myvec{x}_i \in C_1$) and making a few 
quick local adjustments based on $r_i$ --- equation 
(\ref{eq:distanceC1-C0}).

\subsection{A mass R\&D product: Darwinian evolution in parallel universes}
\label{sec:pga}

Let us now consider a different problem, the variable selection problem. 
Given $d$ potential predictors, 
which combination is the best for predicting $y$?
Let $\Omega$ be the space of all possible subsets of 
$C = \{x_1, x_2, ..., x_d\}$.
The typical approach is as follows: First, define a proper evaluation 
criterion,
\[
 F(\omega): \Omega \mapsto \mathbb{R}.
\]
Preferably $F$ should be a {\em fair} measure of $\omega \in \Omega$. 
Common examples of $F$ include the Akaike information criterion 
\citep[AIC,][]{aic}, the Bayesian information criterion 
\citep[BIC,][]{bic}, and generalized cross-validation (GCV), to name a 
few. Then, use a search algorithm to find the best $\omega$ which 
optimizes $F(\omega)$.

\subsubsection{Two challenges: computation and criterion}
\label{sec:varsel-challenges}

There are two main challenges. The first one is computation. With $d$ 
potential predictors, the size of $\Omega$ is $|\Omega|=2^d$. This gets 
large very quickly. For example, take $d=100$ and suppose we can evaluate 
a billion ($10^9$) subsets per second. How long will it take us to 
evaluate all of them? The answer is about 40,000 billion years: 
\[
 2^{100} \div 10^9 \div 3600 \div 24 \div 365 \approx 40,000 \times 10^9.
\]
This may seem serious, but it actually is not the problem we shall be 
concerned about here. Everyone must face this problem; there is no way out 
--- just yet. For moderately large $d$, exhaustive search is impossible; 
stepwise or heuristic search algorithms must be used.

The second challenge is more substantial, especially for statisticians, 
and that's the question of what makes a good evaluation criterion, $F$. It 
is well-known that both the AIC and the BIC are problematic in practice. 
Roughly speaking, with finite data, the AIC tends to favor 
subsets that are too large, while the BIC tends to favor ones
that are too small. 
For classic linear models, both the AIC and the BIC have the form:
\[
 F(\omega) = \mbox{goodness-of-fit}(\omega) + \gamma |\omega|,
\]
where $|\omega|$ is the size of $\omega$, or the number of variables 
included. The AIC uses $\gamma = 2$ whereas the BIC uses $\gamma = 
\log(n)$, $n$ being the sample size. Therefore, it appears that $\gamma=2$ 
is too small and $\gamma=\log(n)$ is too big. But if this is the case, 
surely there must be a magic $\gamma$ somewhere in between? So why not 
find out what it is? While this logic is certainly quite natural, it by no 
means implies that the task is easy. 

\iffalse
In fact, the task is notoriously 
difficult. For example, \citet{yuhong-ABIC} pointed out that, while
BIC is consistent in selecting the true model and AIC is minimax-rate 
optimal for estimating the regression function, it is impossible to 
construct a new information criterion to share these strengths ---
for any model selection criterion to be consistent, it must behave 
suboptimally for estimating the regression function in terms of minimax 
rate of convergence.
\fi

\subsubsection{Darwinian evolution in parallel universes}

The product from my R\&D division is a very simple yet surprisingly 
effective method for variable selection by using Darwinian evolution in 
parallel universes \citep{pga}.

Here is how the algorithm works in a nutshell. Create a number of parallel 
universes. In each universe, run an evolutionary algorithm using the 
(apparently incorrect) AIC as the objective function for {\em just a few 
generations} --- the evolutionary algorithm is a heuristic stochastic 
search algorithm that mimics Darwin's ``natural selection'' to optimize 
any given objective function \citep[see, e.g.,][]{gen-algo}. Whatever it 
is, there will be a current best solution in each universe when we stop. 
For example, the current best subset in universe 1 may be $\{x_3, x_8, 
x_{10}\}$; in universe 2, it may be $\{x_1, x_3, x_8, x_{15}\}$; in 
universe 3, perhaps $\{x_3, x_5, x_8, x_{11}\}$; and so on. These form an 
ensemble. Now take a majority vote and select those variables that show up 
in significantly more universes than the rest. In the example here, this 
would be $\{x_3, x_8\}$ --- and that's the answer.

\subsubsection{Explanation with a toy example}
\label{sec:pga-example}

\begin{figure}[hptb]
 \centering
 \includegraphics[height=3.75in]{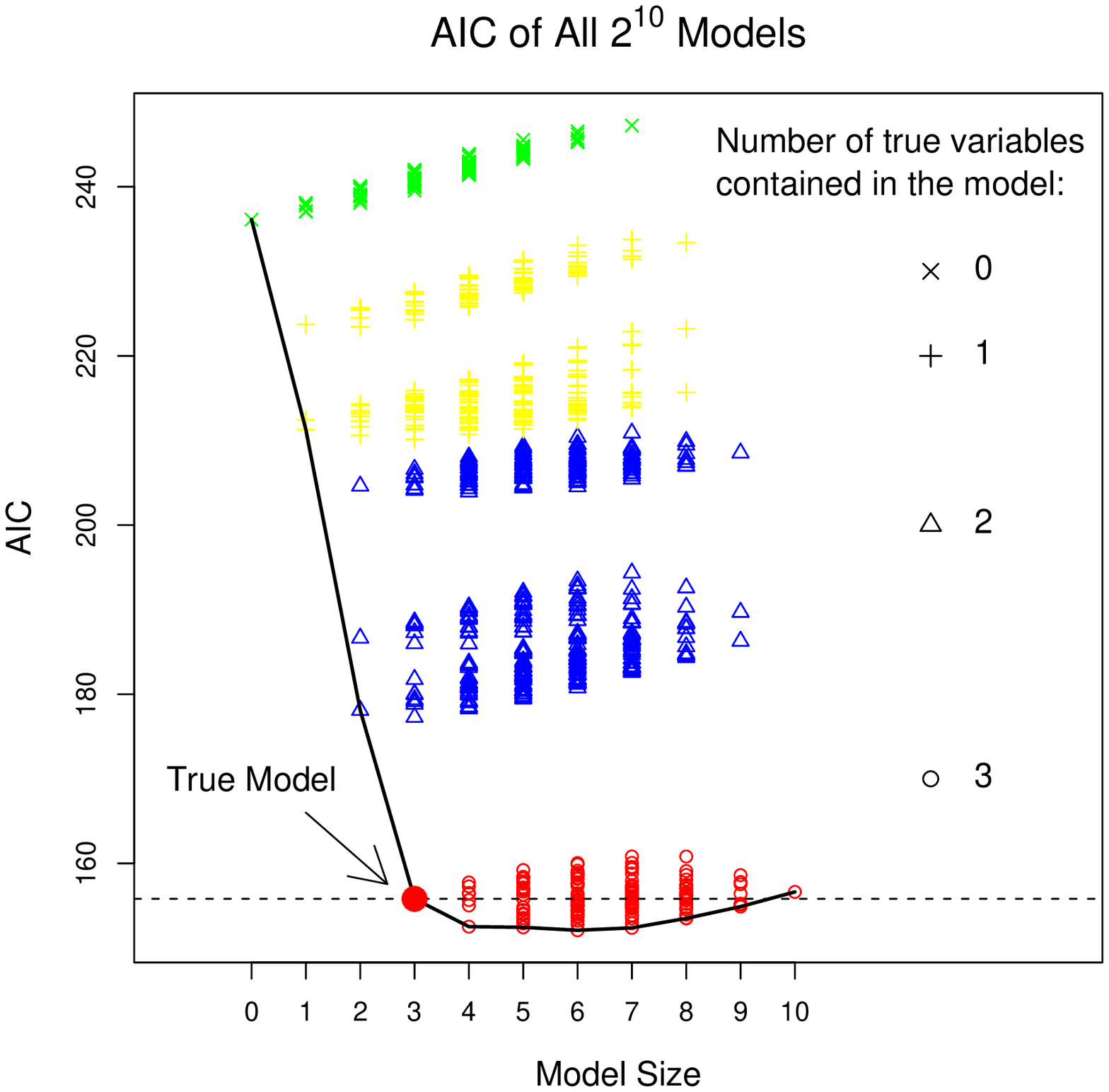}
  \mycap{Why does parallel evolution work? For what this figure tells 
us, see Section~\ref{sec:pga-example}.} 
\label{fig:pga-explain}
\end{figure}

Why does this simple strategy work? A small toy example is enough to 
illustrate the gist of the idea.
Generate 
\[
 y_{i} = x_{i,2} + x_{i,5} + x_{i,8} + \epsilon_i, 
 \quad 
 x_{i,1}, ..., x_{i,10}, \epsilon_i \overset{iid}{\sim} N(0,1),
 \quad
 i=1,2,...,50.
\]
In other words, there are 10 potential predictors but the true model 
contains only 3 of them: $x_2, x_5$, and $x_8$. 
With just 10 
variables, there are altogether $2^{10}=1,024$ subsets, and 
we can still afford to exhaustively compute the AIC for each one of them. 
Figure \ref{fig:pga-explain} plots the AIC versus the size for all 
$2^{10}$ possible subsets. A number of characteristic observations can be 
made: 
\benum

\item\label{AIC:wrong} 
The subset that has the smallest AIC is wrong; it
includes a few variables too many.

\item\label{AIC:nondiscr}
On the AIC scale, many subsets are very close to each other and it is hard 
to tell them apart.

\item\label{AIC:gap}
Let's separate the $2^{10}$ subsets into two groups. Group I consists of 
those that include all the true variables --- they are labelled with 
circles ($\circ$) in the plot. Group II consists of those that miss out on 
at least one of the true variables --- they are labelled with crosses 
($\times$), pluses ($+$), and triangles ($\triangle$). Then, on the AIC 
scale, a significant gap exists between these two groups.

\eenum
Having made these observations, we are now ready to explain why parallel 
evolution works.
The large gap between group I and 
group II (observation \ref{AIC:gap}) means that members from group I are 
significantly superior and hence easily favored by evolution. Therefore, 
after evolving for just a few generations, the current best subset in each 
universe is likely a member from group I. They are the ones that make 
up our ensemble. What do they have in common? They all include the 3 true 
variables.

But in order for majority vote to be effective in selecting the right 
variables, it is necessary that the true variables are the only thing that 
these ensemble members have in common. That's why we can't run the 
evolution for too long in each universe. With a short evolution, since 
members of group I are hard to distinguish from each other on the AIC 
scale (observation \ref{AIC:nondiscr}), the random nature of evolution 
will cause each universe to settle on different members from this group. 
If, on the other hand, we run the evolution for too long, the current best 
subsets from different universes will start to develop something else in 
common --- they will all start to converge to the minimum AIC solution, 
which includes spurious variables (observation \ref{AIC:wrong}).

\begin{figure}[hptb]
 \centering
 \includegraphics[height=1.95in]{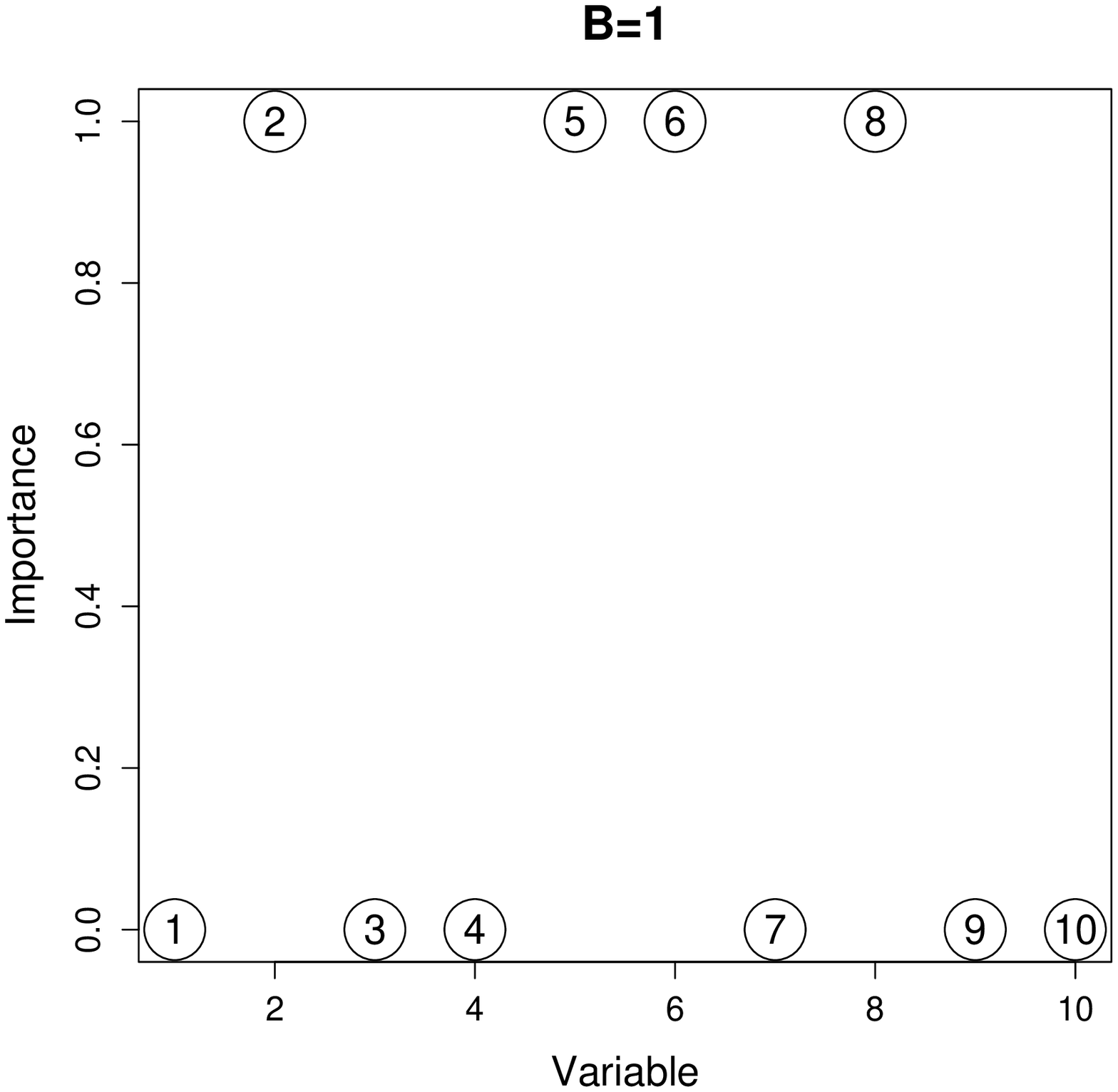}
 \includegraphics[height=1.95in]{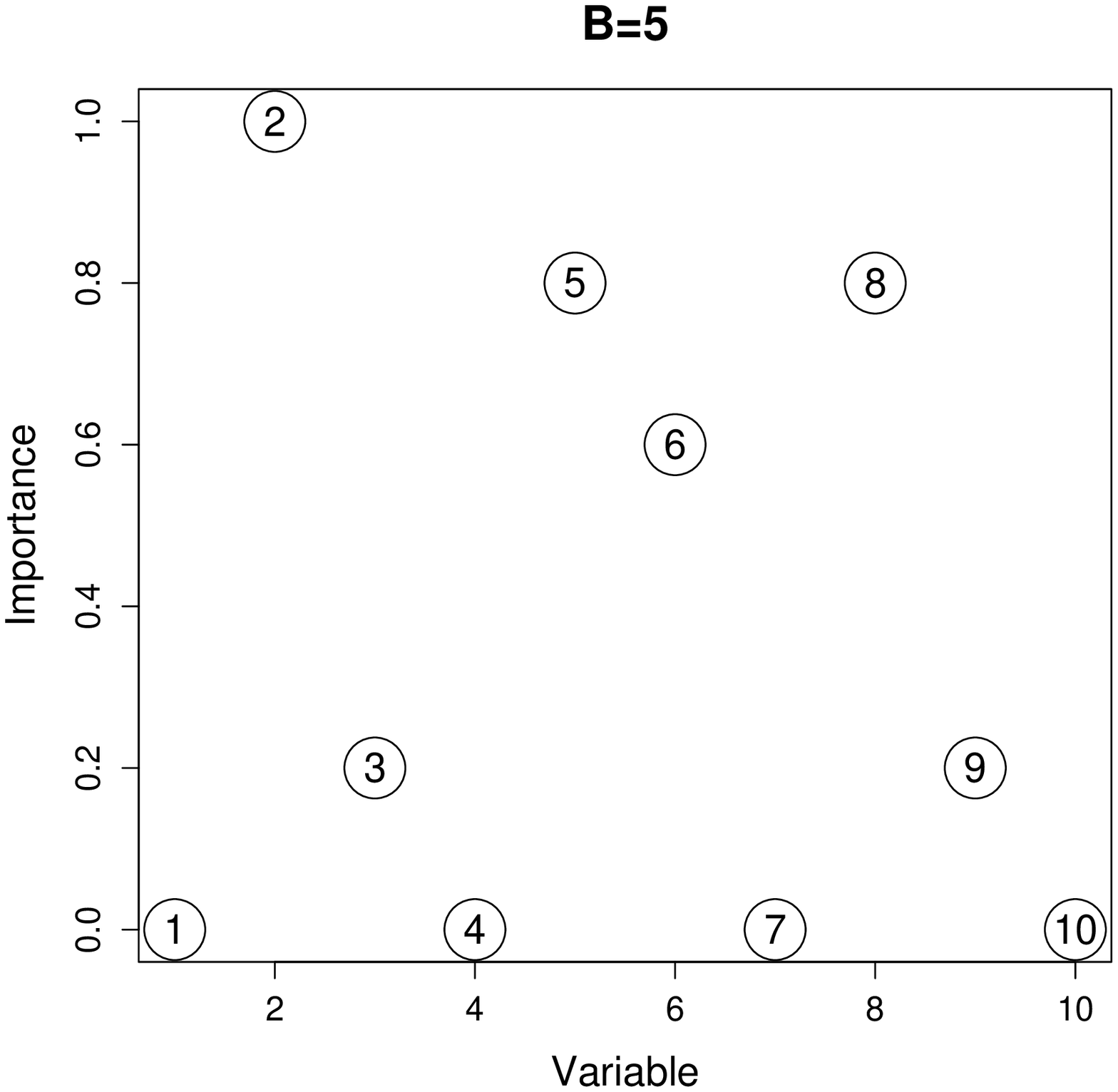}
 \includegraphics[height=1.95in]{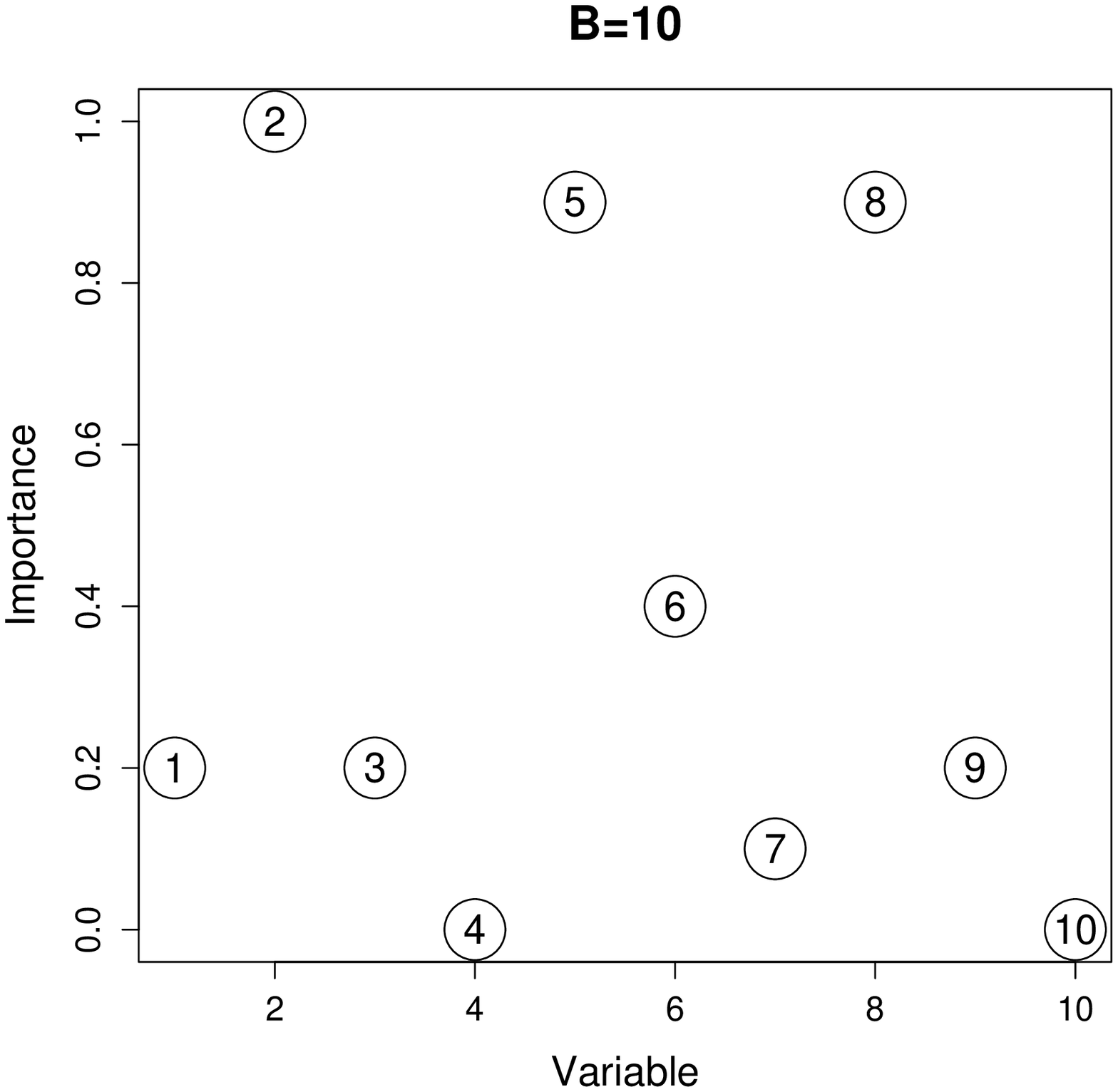}
  \mycap{Parallel evolution on the toy example (Section 
\ref{sec:pga-example}). The correct solution for this example is
$\{2,5,8\}$. When $B=10$ parallel 
universes are used, the correct solution is already easily discernible. }
\label{fig:pga-example}
\end{figure}

Figure \ref{fig:pga-example} illustrates how parallel evolution works on 
this toy example. After running the evolutionary algorithm for just $6$ 
generations in each universe, we measure the importance of a variable by 
how often it shows up across the parallel universes. The correct solution 
for this example is $\{2, 5, 8\}$. When a single universe is used ($B=1$), 
we get the wrong solution --- a spurious variable, namely variable $6$, 
also shows up. But as more and more parallel universes are used, only the 
truly important variables, i.e., variables 2, 5 and 8 in this case, can 
``survive'' the majority vote. We can see from Figure 
\ref{fig:pga-example} that when as few as $B=10$ universes are used, the 
correct solution is already easily discernible: out of the 10 universes, 
variables 2, 5, and 8 each showed up at least 9 times; variable 6 showed 
up 4 times; and all other variables showed up at most twice.

\subsubsection{Discussion: Parallel evolution is an easy-to-use ensemble 
method}

Parallel evolution for variable selection is a successful example of using 
ensembles in a very different context. By using an ensemble, we can 
significantly ``boost up'' the performance of an apparently wrong variable 
selection criterion such as the AIC. The procedure is very easy to use. 
Most importantly, it is {\em trivial} to adapt this principle to general 
variable selection problems regardless of whether the underlying model is 
a classic linear model, a generalized linear model, a generalized additive 
model, a Cox proportional hazard model, or any other model for which the 
question of variable selection is meaningful. As such, it is not unfair to 
call parallel evolution a first-generation, foolproof, off-the-shelf 
variable selector.

A number of smart statisticians have questioned whether it is necessary to 
use the evolutionary algorithm. For example, one can apply Breiman's 
Bagging principle and create an ensemble as follows: Draw a bootstrap 
sample of the data. Using the bootstrap sample, run a stepwise algorithm 
to optimize the AIC and choose a subset. Do this many times, and we get an 
ensemble of subsets. Take majority vote. Clearly, this would also work. I 
have experimented with this idea and found that it is not as effective; 
the probability of selecting the right subset of variables decreases 
significantly in simulation experiments. Why? Breiman's theorem (Section 
\ref{sec:BreimanThm}) points us to an answer. Because bootstrapping alone 
does not create enough diversity within the ensemble. These subsets share 
too many things in common with the minimal AIC solution.

\subsection{Section Summary}
\label{sec:summary-mywork}

In this section, I have discussed a new kernel-based algorithm for rare 
target detection, LAGO, and a new ensemble method for variable selection 
based on parallel evolution. In doing so, a more general formulation of 
LAGO is presented (Section \ref{sec:lago}) using much better mathematical 
notation, e.g., equation (\ref{eq:LAGO}). A simpler version, sLAGO, is 
given for the first time. Better explanations are also given for why 
parallel evolution (Section \ref{sec:pga}) works, e.g., Figure 
\ref{fig:pga-explain}. Many people have the incorrect understanding that 
parallel evolution is merely a better search algorithm for variable 
selection. This is simply not true. In Section \ref{sec:pga}, it is 
emphasized that, instead of a better search {\em algorithm}, parallel 
evolution is actually an ensemble method that boosts up the performance of 
an apparently incorrect search {\em criterion} such as the AIC.

\section{Conclusion}

So, what have we learned? First of all, we learned that, by using kernel 
functions, we can use many linear algorithms such as separating 
hyperplanes and principal component analysis to find nonlinear patterns 
(Section \ref{sec:kerns}). This easily can be done as long as the 
underlying linear algorithm can be shown to depend on the data only 
through pairwise inner-products, i.e., $\myvec{x}^T_i \myvec{x}_j$. Then, 
we simply can replace the inner-product $\myvec{x}^T_i \myvec{x}_j$ with a 
kernel function $K_h(\myvec{x}_i; \myvec{x}_j)$. However, even though such 
a framework is straightforward, we also learned that it is important in 
practice to use the right kernel function $K_h$ and to carefully select 
the hyperparameter $h$ (Section \ref{sec:kernsdisc}). We saw that this is 
not necessarily an easy task (Section \ref{sec:SVMspam}).

We then learned about ensemble methods (Section \ref{sec:ens}). The 
fundamental idea there is to use a collection of perhaps 
not-so-well-tuned models rather than a single model that often 
requires careful fine-tuning. This usually makes ensemble methods easier 
to use for non-experts. I then emphasized that, even for ensembles, it is 
necessary to perform some fine-tuning (Section \ref{sec:ensdisc}) --- this 
typically involves creating the right amount of diversity in the ensemble 
(Section \ref{sec:secret} and \ref{sec:ensdisc}). However, we saw that 
fine-tuning an ensemble algorithm is often easier than fine-tuning a 
kernel-based algorithm (Section \ref{sec:RFspam}).

I then argued (Section \ref{sec:myopinion}) that kernel methods and 
ensemble methods need to co-exist in practice. In particular, non-experts 
may tend to prefer ensemble methods because they are easier to use, 
whereas experts may tend to prefer kernel methods because they provide 
more flexibility for solving nonstandard and difficult problems (Sections 
\ref{sec:kernsdisc} and \ref{sec:ensdisc}). Hence, it is important for 
researchers in statistical machine learning to advance both 
types of methodology. I then presented some of my own research on both 
fronts: LAGO, a fast kernel machine for rare target detection; and 
Darwinian evolution in parallel universes, an ensemble method for 
variable selection.

\section*{Acknowledgment}

This expository article is based on two workshops (September 2006, Sydney, 
Australia; October 2006, Adelaide, Australia) and a graduate course 
(January-April 2007, University of Waterloo, Canada) that bear the same 
title. The workshops were made possible by an AusCan scholarship from the 
Statistical Societies of Australia and Canada. I am indebted to
Dr.~Peter Ricci of the Australian Taxation Office for initiating these 
workshops. I have received many valuable comments and feedbacks from the 
participants of the workshops as well as from the students and faculty 
members who painstakingly sat through my graduate course.
I thank Dr.~Stanley S.~Young for providing me with temporary office space 
at the National Institute of Statistical Sciences (NISS), where this 
article was first written. I also thank Dr.~Dianna Xu and Dr.~Dylan Small 
for providing me with temporary office spaces at Bryn Mawr College and the 
University of Pennsylvania, respectively, so that I could work on the 
revision of this article with little distraction.
My research is partially supported by the Natural Science and Engineering 
Research Council (NSERC) of Canada, Canada's National Program on Complex 
Data Structures (NPCDS) and the Mathematics of Information Technology And 
Complex Systems (MITACS) network. I'd like to acknowledge the 
work by Alexandra Laflamme-Sanders and Dandi Qiao to make LAGO 
and parallel evolution into R libraries; these libraries 
will be publically available from http://cran.r-project.org/.
Finally, I'd like to thank the anonymous referees for helping me improve 
this manuscript, and especially referee number one for giving me the 
strongest support and encouragement possible.

\bibliographystyle{natbib}
\bibliography{TAS-MS07-045RR}

\end{document}